# Scaling of transverse nuclear magnetic relaxation due to magnetic nanoparticle aggregation

K. A. Brown

Harvard School of Engineering and Applied Science, Cambridge, Massachusetts 02138

C. C. Vassiliou

Massachusetts Institute of Technology Department of Electrical Engineering and Computer Science, Cambridge, Massachusetts 02139

D. Issadore[*] and J. Berezovsky

Harvard School of Engineering and Applied Science, Cambridge, Massachusetts 02138

M. J. Cima

Massachusetts Institute of Technology Department of Materials Science and Engineering and Koch Institute for Integrative Cancer Research, Cambridge, Massachusetts 02139

R. M. Westervelt[†]

Harvard School of Engineering and Applied Science and Department of Physics, Cambridge, Massachusetts 02138

**Abstract**

The aggregation of superparamagnetic iron oxide (SPIO) nanoparticles decreases the transverse nuclear magnetic resonance (NMR) relaxation time $T_2^{CP}$ of adjacent water molecules measured by a Carr-Purcell-Meiboom-Gill (CPMG) pulse-echo sequence. This effect is commonly used to measure the concentrations of a variety of small molecules. We perform extensive Monte Carlo simulations of water diffusing around SPIO nanoparticle aggregates to determine the relationship between $T_2^{CP}$ and details of the aggregate. We find that in the motional averaging regime $T_2^{CP}$ scales as a power law with the number $N$ of nanoparticles in an aggregate. The specific scaling is dependent on the fractal dimension $d$ of the aggregates. We find $T_2^{CP} \propto N^{-0.44}$ for aggregates with $d = 2.2$, a value typical of diffusion limited aggregation. We also find that in two-nanoparticle systems, $T_2^{CP}$ is strongly dependent on the orientation of the two nanoparticles relative to the external magnetic field, which implies that it may be possible to sense the orientation of a two-nanoparticle aggregate. To optimize the sensitivity of SPIO nanoparticle sensors, we propose that it is best to have aggregates with few nanoparticles, close together, measured with long pulse-echo times.

## I. INTRODUCTION

Superparamagnetic iron oxide (SPIO) nanoparticles have been used as nuclear magnetic resonance (NMR) chemical sensors by functionalizing them to aggregate in the presence of a specific small molecule.[1] This aggregation decreases the transverse relaxation time $T_2^{CP}$ of the protons in surrounding water that is detectable by a Carr-Purcell-Meiboom-Gill (CPMG) pulse sequence.[1] SPIO nanoparticles have been functionalized to detect the presence of oligonucleotides,[1,2] DNA cleaving agents,[1,2] mRNA,[1,2] enzymes,[1,2] proteins,[1,2] viruses,[1,2] calcium,[3] stereoisomers,[4] and block-copolymers.[5] The technique is well suited for *in vivo* sensing because of its bio-compatibility and because it can be readout by magnetic resonance imaging (MRI).[6] The decrease in $T_2^{CP}$ has been explained theoretically for single nanoparticle systems by inner- and outer-shell theory,[7] then refined with chemical exchange[7,8] and partial refocusing models.[9] The current theoretical understanding of NMR relaxation for SPIO aggregates is based on extending the theory of single nanoparticle systems and treating an aggregate as a single magnetic sphere of larger size.[10,11] Monte Carlo simulations have been performed previously to investigate aggregates with up to 15 nanoparticles by varying particle size and average spacing.[12] These simulations reveal that the effect of aggregation on $T_2^{CP}$ depends on nanoparticle size and that $T_2^{CP}$ only decreases for the aggregation of small nanoparticles.

In this work, we perform in-depth Monte Carlo simulations to investigate the dependence of $T_2^{CP}$ on the detailed characteristics of nanoparticle aggregates. We begin by presenting a summary of the theory pertaining to NMR of single nanoparticles. We then describe the random walk simulation used to calculate $T_2^{CP}$ for nanoparticle aggregate systems. The random walk simulation is verified by showing agreement with theoretical results for single nanoparticles. Simulations of two-nanoparticles show that $T_2^{CP}$ depends strongly on the separation of the nanoparticles and the orientation of the nanoparticles relative to the external magnetic field, an effect that could be used to detect the orientation of nanoparticles aggregates. For larger aggregates, we find that $T_2^{CP}$ scales as a power law with the number $N$ of nanoparticles in an aggregate when the aggregate is in the motional averaging regime. The scaling depends on the fractal dimension $d$ of the aggregate; we find $T_2^{CP} \propto N^{-0.44}$ for aggregates with $d = 2.2$, a typical value for diffusion limited aggregation. We find the power law scaling from the random walk simulations to be reasonable in comparison with a simple model. Based on our observations, we propose that the greatest effect on $T_2^{CP}$ will be caused by aggregates with few nanoparticles, that have minimal separation between nanoparticles, and are measured with long pulse echo times.

## II. NMR OF SINGLE NANOPARTICLE SYSTEMS

An NMR measurement is performed by initially polarizing an ensemble of nuclear moments, then observing their relaxation in time. For the ensemble of nuclear moments, we consider the nuclear magnetic moment of the hydrogen atoms of water. An external z-directed magnetic field $\vec{B}_e$ initially polarizes the average moment $\vec{I}$ in the z-direction. An NMR measurement begins with an applied radio frequency (RF) pulse rotating $\vec{I}$ to the x-axis. The average moment will relax back to the z-axis with spin-lattice relaxation time $T_1$ while each individual moment precesses in the x-y plane at the local Larmor frequency $\omega_L = \gamma B(\vec{r})$, where $\gamma$ is the proton gyromagnetic ratio and $B(\vec{r})$ is the magnitude of the z-directed magnetic field at location $\vec{r}$. The magnitude of $\vec{I}$ will decrease as the moments of the protons lose coherence due to spin-spin interactions and the inhomogeneity of the magnetic field. This transverse decay of $\vec{I}$ is characterized by the effective transverse relaxation time $T_2^*$. A Carr-Purcell-Meiboom-Gill (CPMG) pulse sequence[13] may be used to cancel the effect of field inhomogeneity in order to measure the true transverse decay time $T_2$. The CMPG pulse sequence consists of a second RF pulse at time $\tau_{CP}$ after the first RF pulse which rotates each moment 180 degrees about the z-axis, and causes the spins to realign and form a spin echo at time $2\tau_{CP}$. This process is repeated to obtain a spin-echo train whose envelope decays as[14] $\exp\left[-t\left(T_2^{-1}+T_1^{-1}\right)\right]$.

The transverse decay time $T_2^{CP}$ measured by CPMG pulse sequences in systems with diffusion depends on the relationship between the diffusion time $\tau_D$ for a water molecule to move between magnetic field fluctuations and the pulse-echo time $\tau_{CP}$. If $\tau_{CP} \gg \tau_D$, water will diffuse far compared to variations in the magnetic field between subsequent pulse-echo pairs making the pulse sequence unable to effectively refocus the moments. This is called the Motional Averaging (MA) regime; in this limit, $T_2^{CP}$ is equal to $T_2^*$ and will remain dominated by the field inhomogeneity. If $\tau_{CP} \ll \tau_D$, water will not diffuse far in between pulse-echo pairs, and thus a CPMG pulse sequence will effectively refocus the magnetic moments and $T_2^{CP}$ will approach $T_2$. This is called the Echo Limited (EL) regime.[13]

There is a good theoretical understanding of the effect of single nanoparticles on the transverse decay time $T_2^{CP}$ measured by a CMPG pulse sequence. We consider a spherical nanoparticle with radius $r_p$ and magnetic dipole moment $\vec{m}$ pinned in the z-direction by an external field $\vec{B}_e$. The external field is typically sufficiently strong that such that interactions between nanoparticles can be ignored.[15] The z-component of the magnetic field generated by each nanoparticle is,

$$B_z(r,\theta) = \frac{\mu_0 m}{4\pi}\left(\frac{3\cos^2(\theta)-1}{r^3}\right), \tag{1}$$

where $\vec{r}$ is the position relative to center of the nanoparticle, $\theta$ is the angle to the z-axis, and $\mu_0$ is the permeability of free space. The effect of the nanoparticle is characterized by the root-mean-square variation $\Delta\omega$ in the Larmor frequency on the surface of the nanoparticle and the volume fraction $\Phi$ occupied by nanoparticles. The diffusion time is given by $\tau_D = r_P^2/D$, where $D$ is the self-diffusion constant of water and $r_P$ is the radius of the nanoparticle. In the EL regime ($\tau_{CP} \ll \tau_D$), the transverse relaxation time $T_2^{EL}$ due to the nanoparticle is given by,[7]

$$\frac{1}{T_2^{EL}} = 2.25\, \Phi (\Delta\omega)^2 \frac{\tau_{CP}^2}{\tau_D}. \qquad (2)$$

In the MA regime ($\tau_{CP} \gg \tau_D$), the transverse relaxation time $T_2^{MA}$ due to the nanoparticle is given by,[7]

$$\frac{1}{T_2^{MA}} = \frac{4}{9} \Phi (\Delta\omega)^2 \tau_D. \qquad (3)$$

These equations are valid for the weak dephasing regime where $\Delta\omega^{-1} \gg \tau_{CP}$ and $\Delta\omega^{-1} \gg \tau_D$, as is typically the case for SPIO nanoparticles. We do not consider decay due to $T_1$, typically[16] $T_1 \gg T_2^{CP}$. We set the true transverse decay time $T_2 = \infty$ in order to focus on the effects of the inhomogeneous field.[16] A CPMG pulse sequence is experimentally necessary to cancel relaxation due to inhomogeneous field effects with longer length scales.

### III. METHODS: RANDOM WALK SIMULATION

We investigate the dephasing caused by nanoparticle aggregates using a Monte Carlo method to simulate $T_2^{CP}$ for water molecules diffusing in the magnetic field profile created by a nanoparticle aggregate. Monte Carlo methods for calculating $T_2^{CP}$ are well established.[9,12,17] We begin by constructing a nanoparticle aggregate with diffusion limited aggregation (DLA). The aggregate is then placed in a spherical volume in which water molecules are allowed to diffuse. Calculating diffusive trajectories of water allows us to determine the magnetic field experienced by each proton and thus the dynamics of each magnetic moment. An ensemble of such random walks allows us to calculate the average magnetic moment $\vec{I}$ of the ensemble and observe its decay.

We construct aggregates of $N$ nanoparticles by diffusion limited aggregation (DLA) as shown in Fig. 1(a).[18] A seed nanoparticle is placed at the origin and additional nanoparticles execute random walks until they come into contact with a stationary nanoparticle, at which point they irreversibly stick. The size of an aggregate is approximated by its radius of gyration $r_G$, the root-mean-square distance of the nanoparticles to the center of mass. The fractal dimension $d$ of such aggregates is found to be $d = 2.2 \pm 0.1$ by fitting the aggregates to $r_G \propto N^{1/d}$ over the range $20 < N < 125$. The fractal dimension of aggregates assembled by DLA is typically $d = 2.45 \pm 0.10$ and depends on the details of the DLA technique.[18]

To simulate water diffusing in nanoparticle systems, we calculate an ensemble of random walk trajectories for water molecules. Water is allowed to diffuse in a spherical volume of radius $r_B$ surrounding the nanoparticle aggregate as shown in Fig. 1(b). The bounding radius $r_B = r_P (N/\Phi)^{1/3}$ is picked to hold the volume fraction $\Phi$ constant. Trajectories of water molecules are simulated as three-dimensional isotropic random walks with constant time step $\Delta t$. The position of water molecule $i$ at time $j$ is denoted $\vec{s}_i^{\,j}$. Walkers that collide with nanoparticles deflect randomly off the surface of the nanoparticle to finish their step. Walkers that would step outside $r_B$ stop at the boundary and are translated by $2r_B$ through the origin placing them on the opposite edge where they finish their step in a random direction.

The decay of the average magnetic moment $\vec{I}$ of the ensemble is found by recording the phase accumulated due to Larmor precession of each water molecule during its random walk. The phase $\phi_i$ of the each random walk $i$ begins at $\phi_i = 0$, representing initial alignment in the x-direction. Walker $i$ accumulates phase $\phi_i = \sum_j \gamma B_z(\vec{s}_i^j) \Delta t$ where $B_z(\vec{r})$ is the magnetic field created by the nanoparticles. A CPMG pulse sequence is simulated by inverting the previously accumulated phase $\phi_i \to -\phi_i$ at each pulse $(t = \tau_{CP}, 3\tau_{CP}, 5\tau_{CP}...)$. The ensemble average x-component of the moments is calculated at each spin-echo ($t = 0, 2\tau_{CP}, 4\tau_{CP}, ...$) as the average cosine of the phases $I_x = \langle \cos(\phi) \rangle_i$. The decay of $I_x$ is fit to $I_x = \exp(-t/T_2^{CP})$ to extract the transverse decay time $T_2^{CP}$ for the simulated CPMG pulse sequence. We do not include spin-spin or spin-lattice relaxation in order to focus on the effect of the field inhomogeneity.

By dividing the measured transverse decay time $T_2^{CP}$ by the decay time $T_2^{MA}$ predicted in the motional averaging regime given in Eq. (3), we define a normalized transverse relaxation time $\tilde{T}_2 = T_2^{CP}/T_2^{MA}$ that is invariant of $m$, $r_P$, $D$, and $\Phi$. For all results presented here, we choose conditions that are relevant to chemical sensing experiments.[2,19] We simulate 9 nm diameter maghemite spheres with magnetic moments $m = 2.5 \times 10^4 \mu_B$ where $\mu_B$ is the Bohr magneton. The nanoparticles are chemically coated such that $r_P = 30$ nm and suspended at volume fraction $\Phi = 0.001$ in water with a self-diffusion constant $D = 2.3 \times 10^{-9} \, m^2/s$. These constants define $\tau_D = 390$ ns, $\Delta\omega = 2 \times 10^5$ rad/s, and $T_2^{MA} = 137 \, ms$. Simulations were performed with MATLAB (MathWorks) on the Harvard Odyssey 1024-processor Linux computing cluster.

## IV. RESULTS

The random walk simulation shows exponential decay of the average x-component $I_x$ of the proton magnetic moments in nanoparticle systems. Figure 1(c) shows the decay of $I_x$ for a single nanoparticle system and for an aggregate with $N = 3$. The normalized decay rate of the single particle case is found to be $\tilde{T}_2^{-1} = 0.96$ by fitting $I_x$ to an exponential decay. The three-nanoparticle aggregate is found to have a faster decay rate $\tilde{T}_2^{-1} = 1.51$. The decay of $I_x$ is the result of the inhomogeneous field created by the nanoparticles. These simulations were performed in the motional averaging (MA) regime over 3,800 pulse-echo cycles with $\tau_{CP} = 100\tau_D$, 50,000 diffusing water molecules in the case of $N = 1$, and 125,000 diffusing water molecules in the case of $N = 3$. To minimize computation time, $\tilde{T}_2$ is calculated after one CPMG pulse-echo cycle in subsequent simulations described below.

The normalized decay rate $\tilde{T}_2^{-1}$ is simulated *vs.* the pulse-echo time $\tau_{CP}$ to show that the random walk simulation agrees with the theoretical values of $T_2^{CP}$ given in Eqs. (2) and (3). A plot of the decay rate $\tilde{T}_2^{-1}$ *vs.* $\tau_D/\tau_{CP}$ is shown in Fig. 1 (d) for a single nanoparticle and for an aggregate with $N = 3$. The asymptotic values of $\tilde{T}_2^{-1}$ for $N = 1$ agree with the theoretical values of $T_2^{CP}$ for single nanoparticle systems given in Eqs. (2) and (3), which are shown as grey lines in Fig. 1(d). The simulations exhibit a smooth transition from the MA regime ($\tau_{CP} \gg \tau_D$) to the EL regime ($\tau_{CP} \ll \tau_D$) near $\tau_{CP} = \tau_D$. The decay rate $\tilde{T}_2^{-1}$ for an aggregate with $N = 3$ is also shown in Fig. 1 (d). As expected, there is no significant deviation between the $N = 1$ and $N = 3$ cases in the EL regime because the aggregation of additional nanoparticles has little effect on $\tilde{T}_2^{-1}$ when the pulse-echo cycles can cancel the effect of a single nanoparticle. In the MA regime, $\tilde{T}_2^{-1}$ is 1.6 times larger for $N=3$ than for $N=1$. Each data point in Figure 2(b) represents a simulation done with $2.5 \times 10^5$ to $1 \times 10^7$ diffusing water molecules over 10 to 60,000 time steps.

Simulations of two-nanoparticle aggregate systems show that $\tilde{T}_2^{-1}$ depends strongly on the relative position of the two nanoparticles. Figure 2(a) shows the simulated dependence of $\tilde{T}_2^{-1}$ on the separation $l$ of the nanoparticles and the angle $\theta$ between the axis joining the nanoparticles and the external magnetic field $\vec{B}_e$. Figure 2(b) shows $\tilde{T}_2^{-1}$ averaged over the full $4\pi$ solid angle to obtain $\langle \tilde{T}_2^{-1} \rangle_\theta$ at a given separation, as would be the case in a thermally averaged ensemble. We find that $\langle \tilde{T}_2^{-1} \rangle_\theta$ monotonically increases as two nanoparticles are brought together, demonstrating the mechanism of aggregation as a sensor. The inset of Fig. 2(b) shows $\tilde{T}_2^{-1}$ vs. $\cos(\theta)$ at $l = 2r_P$. This strong angular dependence suggests that it may be possible to sense the orientation of a two-particle aggregate. Figure 2(a) represents simulations of 4,000 aggregates with random values of $l$ and $\theta$ with $2.5 \times 10^5$ diffusing water molecules taking $2 \times 10^5$ steps each in the MA regime with $\tau_{CP} = 100 \tau_D$.

Figure 3(a) depicts $\tilde{T}_2^{-1}$ vs. $N$ in the motional averaging (MA) regime with each point representing an individual aggregate. These results show that $\tilde{T}_2^{-1}$ scales as a power law with the number $N$ of nanoparticles in an aggregate. For aggregates assembled by diffusion limited aggregation with a fractal dimension of $d = 2.2$, we find the scaling to be $\tilde{T}_2^{-1} \propto N^{0.44}$. The grey line in Fig. 3(a) represents a fit to $\tilde{T}_2^{-1} = bN^\alpha$ which gives $\alpha = 0.444 \pm 0.002$. The spread in points about the fit is due to the dependence of $\tilde{T}_2^{-1}$ on the geometry of each aggregate. This scaling relationship implies that NMR measurements can yield quantitative information about the number of nanoparticles in an aggregate. The fact that $\alpha < 1$ indicates that the marginal change in $\tilde{T}_2^{-1}$ decreases as each additional nanoparticle is added to an aggregate. Therefore, smaller aggregates are more effective as sensors of aggregation. Figure 3(a) represents simulations of 500 individual aggregates, each with 5,000 diffusing water molecules taking $3 \times 10^6$ steps each in the MA regime with $\tau_{CP} = 5,000 \tau_D$.

The scaling of $\tilde{T}_2^{-1}$ with $N$ is independent of pulse-echo time $\tau_{CP}$ when aggregates are in the MA regime. Figure 3(b) shows the scaling exponent $\alpha$ vs. $\tau_D / \tau_{CP}$, found for aggregates with $N = 25$ by solving $\tilde{T}_2^{-1}\big|_{N=25} = \tilde{T}_2^{-1}\big|_{N=1} 25^\alpha$ for $\alpha$. At $\tau_D / \tau_{CP} < 10^{-3}$, $\alpha$ has saturated to an average value of $\alpha = 0.444 \pm 0.006$, indicated in Fig. 3(b) by the grey line. The saturation of $\alpha$ is equivalent to the MA regime saturation of $\tilde{T}_2^{-1}$ shown in Fig. 1(d). The transition from the MA regime to the EL regime for an aggregate will occur near $\tau_{DG} / \tau_{CP} = 1$ where $\tau_{DG} = r_G^2 / D$ is the diffusion time of water near an aggregate. The radius of gyration $r_G$ of aggregates with $N = 25$ is found to be $r_G \approx 4.7 r_P$ which implies that the transition between the MA and EL regime for aggregates with $N = 25$ should occur near $\tau_D / \tau_{CP} = 5 \times 10^{-2}$. Even though the aggregate has not transitioned into the EL regime, $\alpha$ has decreased from a saturated value of 0.44 to 0.33 at $\tau_D / \tau_{CP} = 1 \times 10^{-2}$, similar to the decrease in $\tilde{T}_2^{-1}$ in the corresponding range $\tau_D / \tau_{CP} = 2 \times 10^{-3}$ to $\tau_D / \tau_{CP} = 2 \times 10^{-1}$ observed in Fig. 1(d). Each point in Fig. 3(b) is the average of 50 distinct aggregate geometries simulated with between $1 \times 10^3$ and $1 \times 10^6$ random walks of between $6 \times 10^5$ and $6 \times 10^6$ steps.

## V. DISCUSSION

A simple explanation for the power law scaling of $\tilde{T}_2^{-1}$ with $N$ is found by modeling a nanoparticle aggregate as a single sphere with radius and magnetic moment larger than those of a single nanoparticle. Following the analysis of Shapiro *et. al.*,[11] an aggregate with fractal dimension $d$ is modeled as a sphere with radius equal to the aggregate's radius of gyration $r_G \propto r_P N^{1/d}$ and magnetic moment $= mN$ because the nanoparticle moments are aligned by the external magnetic field, as in the high-field limit of Langevin's Law.[8] Holding the volume fraction $\Phi$ fixed as $N$ is varied, this scaling implies $\tau_D \propto N^{2/d}$ and $\Delta\omega \propto N^{1-3/d}$, which gives $\tilde{T}_2^{-1} \propto N^{2-4/d}$ when combined with Eq. (3). Using $d = 2.2$ as in our simulation, this predicts the scaling $\tilde{T}_2^{-1} \propto N^{0.18}$. Holding $\Phi$ fixed might not be justified, as the volume occupied by the aggregate is different from the volume of the equivalent sphere if $d \neq 3$. If instead the radius $r_B$ of the simulation volume is held fixed, the volume fraction scales as $\Phi \propto N^{3/df-1}$ which specifies $\tilde{T}_2^{-1} \propto N^{1-1/df}$ and $\tilde{T}_2^{-1} \propto N^{0.55}$. In summary, modeling the aggregate as a sphere predicts power law scaling of $\tilde{T}_2^{-1}$ with $N$ with an exponent between 0.18 and 0.55, which agrees with the exponent $\alpha = 0.44$ from our random walk simulations of nanoparticle aggregates.

The predicted dependence of $\tilde{T}_2^{-1}$ on $N$ and the relative position of nanoparticles in an aggregate could be explored in NMR studies. The strong dependence of $\tilde{T}_2^{-1}$ on the relative position of nanoparticles in an aggregate, as shown in Fig. 2(a), suggests that it may be possible to measure the orientation of two-nanoparticle aggregates. This could be tested by initially aligning aggregates in a high magnetic field, immobilizing them on a surface or in a polymer, then measuring $T_2^{CP}$ while varying the angle between the external NMR field and the alignment of the sample. The scaling relationship $\tilde{T}_2^{-1} \propto N^\alpha$ could be measured if aggregates with precise $N$ could be constructed and $T_2^{CP}$ could be compared with $r_G$ which could be measured by a technique such as dynamic light scattering.[20] If $\alpha$ could be measured for a given system, this scaling could be used to extract quantitative information from NMR studies such as the average number of nanoparticles in aggregates.

The results presented here provide three ways to optimize the sensitivity of a SPIO NMR chemical sensor. (1) In a two-nanoparticle system, $\tilde{T}_2^{-1}$ monotonically increases as the two particles are brought together, meaning that nanoparticles in an aggregate should be as close as to each other possible. (2) As shown in Fig. 3(b), the decrease in the exponent $\alpha$ for decreasing values of $\tau_{CP}$ means that the change in $\tilde{T}_2^{-1}$ due to aggregation is greatest when $\tau_{CP}$ is much longer than $\tau_D$, ensuring operation in the MA regime. (3) The value of $\alpha$ is predicted to depend strongly on the details of aggregation, but the fact that $\alpha < 1$ is important to sensor design. The shift in $T_2$ due to adding a single nanoparticle to an aggregate becomes lower as $N$ increases because $\alpha < 1$. For a constant number of nanoparticles and a limited number of binding agents, nanoparticles with fewer binding sites are preferable as sensors.

## Acknowledgements

We would like to acknowledge Mike Stopa for helpful discussions. We would also like to acknowledge generous support by the Department of Defense through the National Defense Science & Engineering Graduate Fellowship (NDSEG) Program, the Harvard-MIT Center for

References:

[*]Present address: Massachusetts General Hospital Center for Systems Biology, Boston, Massachusetts 02114

[†]Electronic address: westervelt@seas.harvard.edu.

1. J. M. Perez and C. Kaittanis, in *Nanoparticles in Biomedical Imaging*, edited by J. W. M. Bulte and M. J. Modo (Springer, New York, 2008) Chap. 9.

2. J. M. Perez, L. Josephson, and R. Weissleder, Chem. Bio. Chem. **5,** 3 (2004).

3. T. Atanasijevic, M. Shusteff, P. Fam, and A. Jasanoff, Proc. Natl. Acad. Sci. **103**, 40 (2006).

4. A. Tsourkas, O. Hofstetter, H. Hofstetter, R. Weissleder, and L. Josephson, Angew. Chem. Int. Ed. **43,** 18 (2004).

5. J.-F. Berret, N. Schonbeck, F. Gazeau, D. E. Kharrat, O. Sandre, A. Vacher, and M. Airiau, J. Am. Chem. Soc. **128,** 5 (2006).

6. K. D. Daniel, G. Y. Kim, C. C. Vassiliou, F. Jalali-Yazdi, R. Langer, and M. J. Cima, Lab Chip **7,** 10 (2007).

7. R. A. Brooks, F. Moiny, and P. Gillis, Magn. Reson. Med. **45,** 6 (2001).

8. R. A. Brooks, Magn. Reson. Med. **47,** 2 (2002).

9. P. Gillis, F. Moiny, and R. A. Brooks, Magn. Reson. Med. **47,** 2 (2002).

10. A. Roch, Y. Gossuin, R. N. Muller, and P. Gillis, J. Magn. Magn. Mater. **293,** 1 (2005).

11. M. G. Shapiro, T. Atanasijevic, H. Faas, G. G. Westmeyer, and A. Jasanoff, Magn. Reson. Imaging **24**, 4 (2006).

12. Y. Matsumoto and A. Jasanoff, Magn. Reson. Imaging **26,** 7 (2008).

13. H. Y. Carr and E. M. Purcell, Phys. Rev. **94,** 3 (1954).

14. E. L. Hahn, Phys. Rev. **80**, 4 (1950).

15. S. H. Koenig and K. E. Kellar, Magn. Reson. Med. **34,** 2 (1995).

16. R. M. E. Valckenborg, H.P. Huinink, J. J. v. d. Sande, and K. Kopinga, Phys. Rev. E **65,** 2 (2002).

17. G. Q. Zhang and G. J. Hirasaki, J. Magn. Reson. **163,** 1 (2003).

18. P. Meakin, Phys. Rev. A **29,** 2 (1984).

19. M. F. Hansen and S. Mørup, J. Magn. Magn. Mater. **184,** 3 (1998).

20. M. Y. Lin, H. M. Lindsay, D. A. Weitz, R. C. Ball, R. Klein, and P. Meakin, Proc. R. Soc. Lond. A **423**, 1864 (1989).

FIG. 1. (a) Schematic diagram showing the diffusion limited aggregation (DLA) of superparamagnetic iron oxide (SPIO) nanoparticles. (b) Illustration of a water molecule executing a random walk in a sphere of radius $r_B$. (c) Simulated ensemble averaged x-component $I_x$ of water nuclear magnetic moments *vs.* time $t$ plotted at multiples of $33\tau_{CP}$ for a single particle and a three-nanoparticle aggregate where $\tau_{CP}$ is the Carr-Purcell-Meiboom-Gill pulse-echo delay time. The size of the points is greater than the statistical error. Linear fits show that both curves are exponential decays. (d) Simulated normalized transverse decay rate $\tilde{T}_2^{-1}$ *vs.* $\tau_D/\tau_{CP}$ for a single nanoparticle and for a three-nanoparticle aggregate, compared to single nanoparticle theory, shown as grey lines; here $\tau_D$ is the diffusion time of a water molecule diffusing through the inhomogeneous magnetic field created by nanoparticles. Data point size is greater than statistical error.

FIG. 2. (a) Simulated transverse decay rate $\tilde{T}_2^{-1}$ normalized by the motional averaging decay rate [Eq. (3)] of a two nanoparticle system *vs.* nanoparticle separation $l$ of the nanoparticles and $\cos(\theta)$ where $\theta$ is the angle between the axis joining the nanoparticles and the external magnetic field $\vec{B}_e$, as shown in the inset. (b) $\tilde{T}_2^{-1}$ averaged over the full $(4\pi)$ solid angle to find $\langle \tilde{T}_2^{-1} \rangle_\theta$ at a given $l$. The grey line indicates $\tilde{T}_2^{-1}$ for a single nanoparticle system. The inset shows $\tilde{T}_2^{-1}$ *vs.* $\cos(\theta)$ at $l = 2r_P$, where $r_P$ is the nanoparticle radius.

FIG. 3. (a) Normalized transverse decay rate $\tilde{T}_2^{-1}$ *vs.* number $N$ of nanoparticles in an aggregate. Each point is a simulation of a randomly assembled aggregate conformation. The line is a power

law fit with an exponent of $\alpha = 0.444 \pm 0.002$. Spread about the fit is due to the dependence of $\tilde{T}_2^{-1}$ on the geometry of the aggregate. Error bars denote statistical error in each simulation. (b) Scaling exponent $\alpha$ vs. the ratio $\tau_D/\tau_{CP}$ of the diffusion time $\tau_D$ to pulse-echo time $\tau_{CP}$ calculated by comparing aggregates with $N = 1$ and $N = 25$ at different values of $\tau_{CP}$. The horizontal grey line denotes the average $\alpha = 0.444 \pm 0.006$ for $\tau_D/\tau_{CP} < 10^{-3}$. The data point indicated as an 'x' represents the value of the fit in Fig. 3(a). The decrease in $\alpha$ as $\tau_D/\tau_{CP}$ increases is due to the aggregate transitioning from the motional (MA) regime to the echo limited (EL) regime.

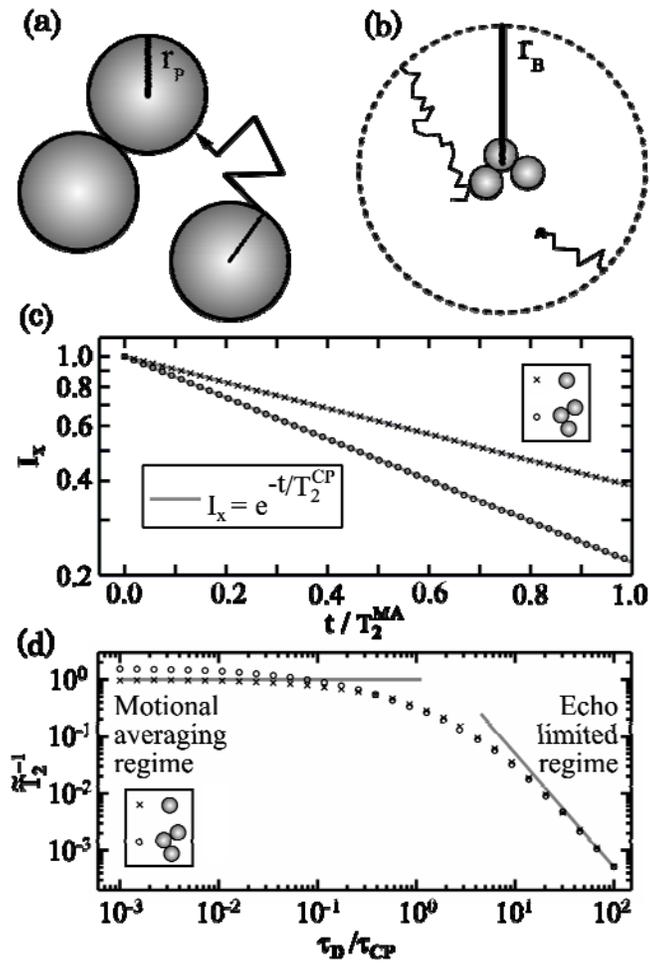

Fig. 1

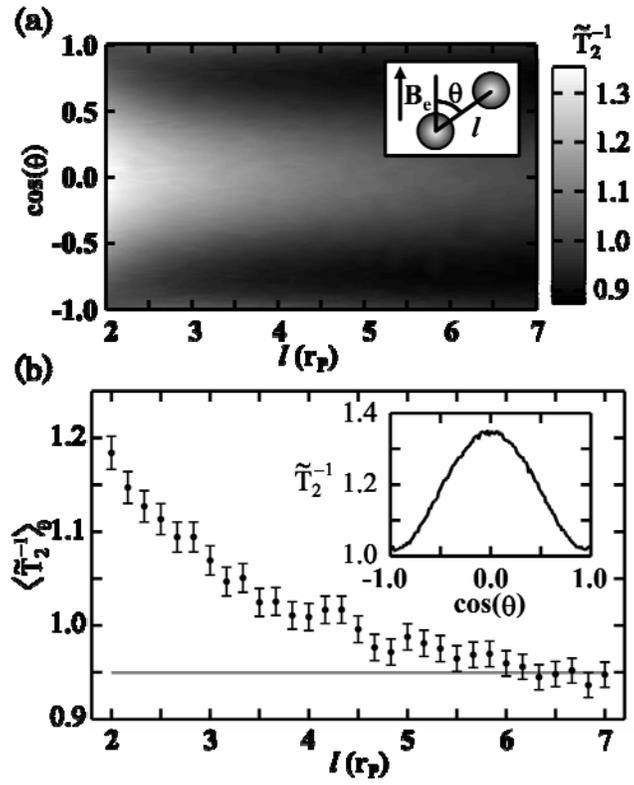

Fig. 2

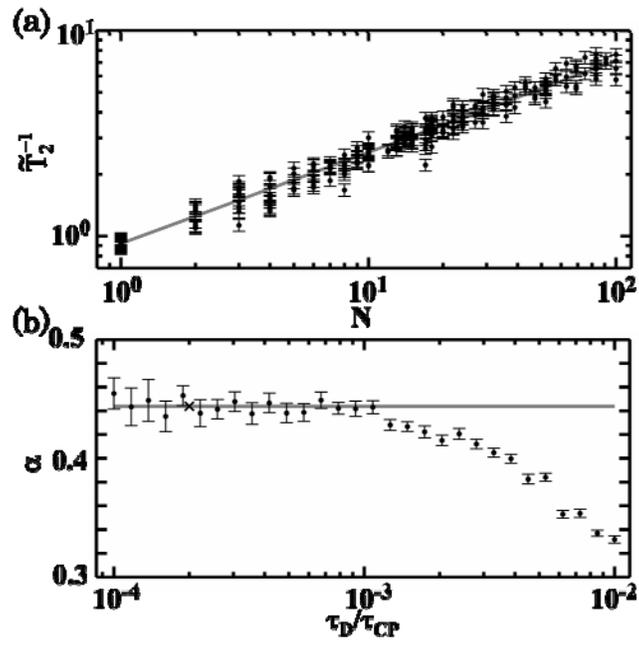

Fig. 3